\documentclass[aps,letterpaper,
showpacs,
twocolumn,prl]{revtex4}
\newcommand{\expec}[1]{\left < #1\right >}

\usepackage{amsmath}
\usepackage{amsfonts}
\usepackage{graphicx}
\usepackage{amsthm}
\usepackage{subfigure}
\usepackage{hyperref}
\usepackage[usenames]{color}
\usepackage{slashbox}

\usepackage{rotating}

\usepackage{braket}

\renewcommand{\v}[1]{\mathbf{#1}}

\newcommand{\lp}{\left ( }
\newcommand{\rp}{\right ) }
\newcommand{\lb}{\left [ }
\newcommand{\rb}{\right ] }

\newcommand{\hc}{\text{H.c.}}

\newcommand{\beq}{\begin{eqnarray*}}
\newcommand{\eeq}{\end{eqnarray*}}

\newcommand{\be}{\begin{eqnarray}}
\newcommand{\ee}{\end{eqnarray}}

\newcommand{\mc}{\mathcal}

\def\lsim{\mathrel{\rlap{\lower4pt\hbox{\hskip1pt$\sim$}}
    \raise1pt\hbox{$<$}}}                

\def\gsim{\mathrel{\rlap{\lower4pt\hbox{\hskip1pt$\sim$}}
    \raise1pt\hbox{$>$}}}                

\begin{document}

\title{Far from equilibrium quantum magnetism with ultracold polar molecules}
\author{Kaden R.~A. Hazzard} \email{kaden.hazzard@colorado.edu}
\affiliation{JILA, NIST,  and Department of Physics, University of Colorado-Boulder, Boulder, Colorado 80309-0440, USA}
\author{Salvatore R. Manmana}
\affiliation{JILA, NIST,  and Department of Physics, University of Colorado-Boulder, Boulder, Colorado 80309-0440, USA}
\author{Michael Foss-Feig}
\affiliation{JILA, NIST,  and Department of Physics, University of Colorado-Boulder, Boulder, Colorado 80309-0440, USA}
\author{Ana Maria Rey}
\affiliation{JILA, NIST,  and Department of Physics, University of Colorado-Boulder, Boulder, Colorado 80309-0440, USA}

\begin{abstract}
Recent theory has indicated how to emulate tunable models of quantum magnetism with ultracold polar molecules.
Here we show that \textit{present} molecule optical lattice experiments can accomplish three crucial goals for quantum emulation,  despite
currently being well below unit filling and \textit{not} quantum degenerate. The first is to verify and benchmark the models proposed to describe these systems. The second is to prepare   correlated and possibly useful states in well-understood regimes.   The third is to explore  many-body physics inaccessible to existing theoretical techniques.
Our proposal relies on a non-equilibrium protocol that can be viewed either as Ramsey spectroscopy or an interaction quench. It uses only routine experimental tools available in any ultracold molecule experiment.
\end{abstract}

\maketitle

Excitement about the recent achievement of near-degenerate ultracold polar molecules~\cite{ni_high_2008,ospelkaus_controlling_2010,ospelkaus:quantum-state_2010,ni:dipolar_2010,miranda:controlling_2011} in optical lattices~\cite{chotia_long-lived_2012}
stems from their strong dipolar interactions and rich internal structure, including rotational, vibrational, and hyperfine states.  These features may be applied to tests of fundamental constants~\cite{zelevinsky:precision_2007}, quantum information~\cite{demille:quantum_2002}, ultracold chemistry~\cite{carr:cold_2009}, and quantum emulation of condensed matter models~\cite{lahaye_physics_2009,trefzger_ultracold_2011,baranov_condensed_2012}. In this paper our focus is on molecules as emulators of quantum magnetism~\cite{barnett:quantum_2006,micheli_toolbox_2006,
buechler:three-body_2007,watanabe:effect_2009,wall:emergent_2009,
yu:spin_2009,krems:cold_2008,wall:hyperfine_2010,
Schachenmayer:dynamical_2010,perez-rios:external_2010,
trefzger:quantum_2010,kestner:prediction_2011}, specifically as proposed in Refs.~\cite{gorshkov:tunable_2011,gorshkov:quantum_2011}.  Models of quantum magnetism have some of the simplest many-body Hamiltonians, yet describe numerous materials~\cite{auerbach:interacting_1994,sachdev_quantum_2008,
lacroix_introduction_2011} and display condensed matter phases ranging from fundamental to exotic: antiferromagnets, valence bond solids, symmetry protected topological phases, and spin liquids.
Emulating quantum magnetism with molecules is appealing because, like cold atoms, the systems are clean and the microscopics well understood.
Advantages over cold atom emulations of quantum magnetism~\cite{bloch_many-body_2008} include orders of magnitude larger energy scales and more tunable Hamiltonians~\cite{gorshkov:tunable_2011,gorshkov:quantum_2011}.
These prior studies have focused on spin \textit{ground states} of unit filling insulators. In contrast, we propose a simple \textit{dynamic} procedure applicable to present experiments, which are ultracold, but non-degenerate and low density.  We show that interesting many body quantum magnetism can be studied immediately.

\begin{figure}[t]
\setlength{\unitlength}{1.0in}
\includegraphics[width=3.5in,angle=0]{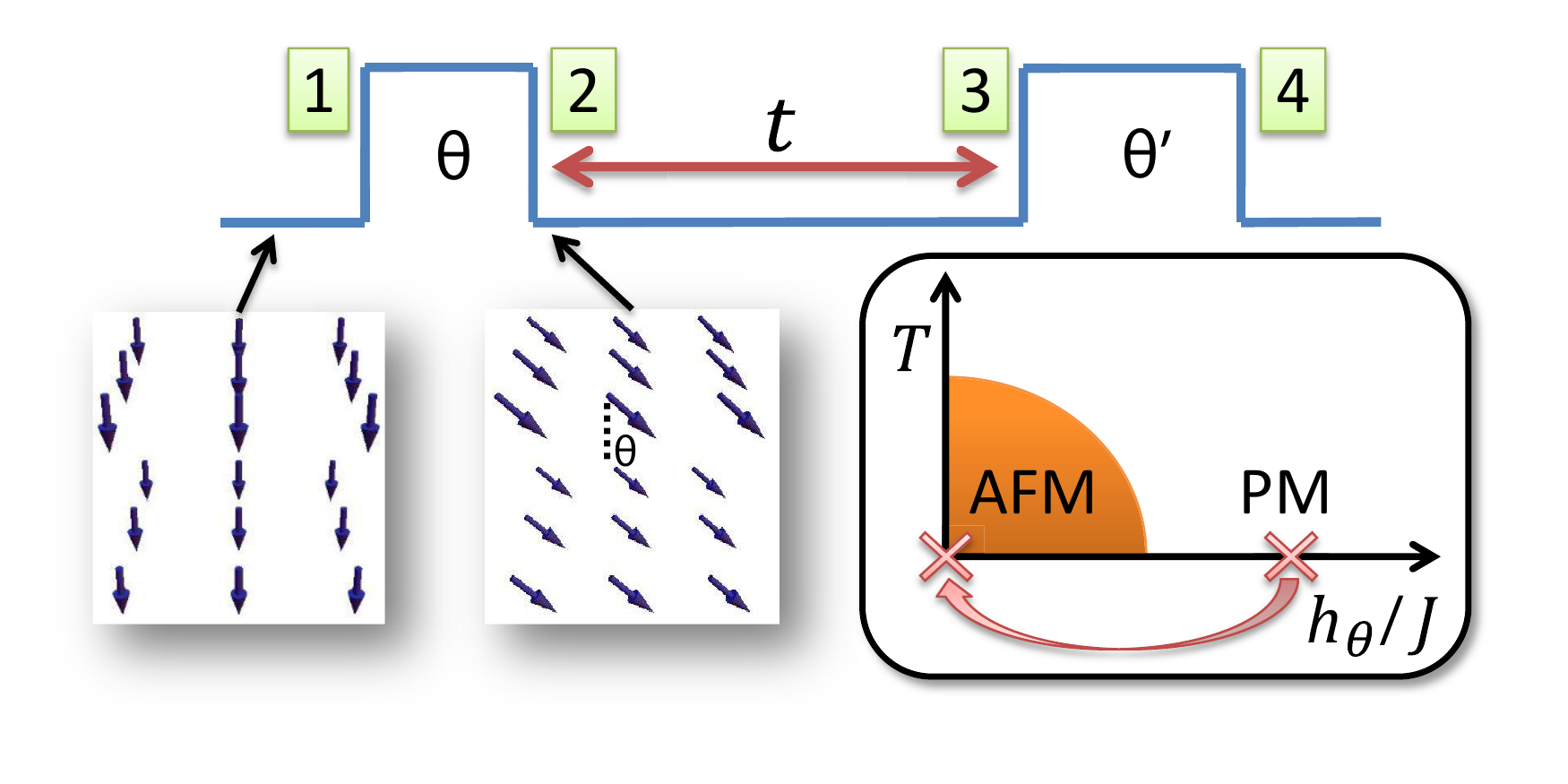}
\vspace{-0.3in}
\caption{ (Color online) Dynamic protocol viewed as Ramsey spectroscopy, two microwave pulses of area $\theta$ and $\theta'$ separated by time $t$. Inset: alternatively viewed as an interaction quench, a sudden $t=0$ turning off of an infinitely strong field $h_\theta$ along $\theta$.
 \label{fig:dynamic protocol}}
\end{figure}

Specifically we  show how experiments may use this dynamics to achieve major goals for emulating quantum magnetism, and we outline these goals to motivate our calculations. First, although
interesting models of quantum magnetism are predicted
to describe
ultracold molecules under appropriate circumstances, this has yet to be experimentally demonstrated.  The proposed dynamic protocol allows such a demonstration as well as benchmarking of the emulator's accuracy.  Second, one wishes to prepare interesting correlated --- and possibly useful --- states.  This protocol can generate
such states in well-understood regimes.  Finally, one wants to explore behavior in these models in regimes inaccessible to present theoretical tools.
This is the generic case for the proposed dynamics.
We emphasize that  all of these goals are achievable under \textit{existing} experimental conditions~\cite{chotia_long-lived_2012}, despite present experiments being \textit{non}-quantum degenerate and at low density. Furthermore, they require only routinely used measurement and preparation tools~\cite{neyenhuis:anisotropic_2012}.

\textit{Background.---}Refs.~\cite{gorshkov:quantum_2011,gorshkov:tunable_2011} show how molecule rotational states  can serve as effective spins, and that dipolar interactions provide an effective spin-spin interaction.  In the simplest case, one populates two rotational levels in a dc electric field $\v{E}$~\footnote{The electric field ensures that the two rotational levels are sufficiently off-resonant from other levels that they form an isolated spin-1/2~\cite{gorshkov:quantum_2011}.} and works in a deep lattice to allow no tunneling.   In this limit, a spin-1/2 dipolar quantum XXZ model describes the molecules~\footnote{We neglect hyperfine coupling, as Refs.~\cite{gorshkov:tunable_2011,gorshkov:quantum_2011} justify.}:
\be
\hspace{-0.1in}H \!&=&\! \frac{1}{2}\sum_{i\ne j}V_{\text{dd}}(i,j)\lb J_z S_i^z S_j^z+\frac{J_\perp}{2}\lp S_i^+ S_j^- + \hc
\rp \rb.\label{eq:XXZ-Ham}
\ee
The sum runs over all molecules, $S_i^z$ and $S_i^\pm$ are the spin-1/2 operators satisfying $[S_i^z,S_i^\pm]=\pm S_i^\pm$, and $V_{\text{dd}}(i,j)=(1-3\cos^2\Theta_{ij})/|\v{r}_i-\v{r}_j|^3$ with $\v{r}_i$ the
$i$'th
molecule's position in lattice units and $\Theta_{ij}$ the angle between $\v{E}$ and $\v{r}_i-\v{r}_j$.
For simplicity and concreteness we assume a dimension $d\le2$ system with $\v{E}$ perpendicular to it, so $V_{\text{dd}}(i,j)=1/|\v{r}_i-\v{r}_j|^3$, but our ideas apply in arbitrary geometries.  One may tune  $J_\perp/J_z$  by changing $\v{E}$ and the choice of rotational state.  We denote by $\ket{0}$, $\ket{1}$, and $\ket{2}$ the three lowest energy rotational eigenstates in the applied $\v{E}$-field with zero angular momentum along the quantization axis. Choosing $\ket{0}$ and $\ket{1}$ to make the spin-1/2, one can tune $\infty>J_\perp/J_z>0.35$ using (readily achievable) $\v{E}$-fields from $0$ to $16$~kV/cm.
 Choosing $\ket{0}$ and $\ket{2}$ for the spin-1/2, one can tune $0<J_\perp/J_z<0.1$ for similar $\v{E}$-fields.  A characteristic scale for these couplings is 400~Hz in KRb and 40~kHz in LiCs~\cite{deiglmayr:formation_2008}, compared to $\lsim$10Hz in cold atoms using superexchange~\cite{trotzky:time-resolved_2008}. 
KRb molecules recently have been loaded in a deep three-dimensional lattice with 25 second lifetimes~\cite{chotia_long-lived_2012},  allowing dynamics lasting thousands of  $J_\perp^{-1}$ and $J_z^{-1}$.

One important aspect of the ongoing experiments is that the filling $f$ is much less than one molecule per site.  The JILA experiments estimate $f\sim 0.1$.  As a result, although the molecules' positions are static throughout one shot, they fluctuate shot-to-shot. Thus, rather than forming a regular lattice, the spins' locations have significant disorder.  Our calculations show that the dynamic protocol's utility persists with disorder.

We use a simple disorder model that likely describes current experiments.  We assume that each site is occupied with a probability $p$ that is independent of other sites~\footnote{In addition to simplicity, this disorder distribution results from suddenly quenching tunneling $t$ to zero for lattice fermions initially at a temperature $T\gg t$.}.
If the molecules are fermions (e.g., KRb~\cite{chotia_long-lived_2012}) then for current temperatures, which occupy only the lowest band, no sites can be doubly occupied and $p=f$.  This also applies to bosons with a strong on-site density-density interaction (e.g., RbCs~\cite{takekoshi:towards_2012}).
The trap causes $f$ to vary spatially.  Although we show results only for the homogeneous system, we have taken
the trap into account
and found that our conclusions remain valid~\cite{hazzard:molecule_in-prep}.

Remarkably, close relatives of such seemingly unusual models exist in the literature, for example the Blume-Emery-Griffiths model~~\cite{blume:ising_1971}.  These mainly focus on the  classical equilibrium limit, $J_\perp=0$.  They were introduced to understand materials~\cite{aeppli:spin_1984,kraemer:dipolar_2012}, $^3$He-$^4$He mixtures~\cite{blume:ising_1971},  Griffiths phases~\cite{griffiths:nonanalytic_1969}, glassy dynamics~\cite{randeria:low-frequency_1985,quilliam:experimental_2012}, and the interplay of the underlying lattice's statistical mechanics (site-dilution percolation) with that of the magnetism living on that lattice~\cite{gefen:anomalous_1983,henley:critical_1985,trinh:correlations_2012}.
Rather than studying
 unique disorder-induced behavior, we focus on showing that
$f=1$ behavior survives
disorder.

Only through  ``disorder" does temperature enter, because the deep lattice freezes out the motion.  In particular, one must distinguish motional temperature from spin temperature.  The former may be large but is entirely captured by the disorder, while the latter is ill-defined since we consider non-equilibrium spin states.
However, experimental microwave manipulation can produce essentially zero entropy spin states.  While one could worry that disorder washes out the behavior,  we will show  that strong correlations, entanglement, and interesting many body physics survive large amounts of disorder.

\textit{Dynamic protocol.---}Our dynamic procedure may be alternatively viewed as Ramsey spectroscopy or an interaction quench (Fig.~\ref{fig:dynamic protocol}). Ref.~\cite{hazzard:spectroscopy_2012} studied closely related Rabi spectroscopy.
In Ramsey spectroscopy, a well established tool in atomic physics, one begins with all molecules in the rotational ground state and applies two strong, resonant  microwave pulses separated by time $t$.   The first pulse initializes the spin states along $\theta$, specifically to $\cos(\theta/2)e^{i\varphi/2} \ket{\downarrow}+\sin(\theta/2) e^{-i\varphi/2}\ket{\uparrow} $, for an angle $\theta$ set by the pulse area, with  high fidelity ($>99\%$).  We take $\varphi=0$ with no loss of generality.  The second pulse rotates a desired spin component, chosen by the pulse area and phase, to the $z$ axis.  In this way one can measure any desired collective spin component $\langle \hat n\cdot \v{S} \rangle$, where $\hat n$ is a unit vector and $S^\alpha=\sum_i S_i^\alpha$ with $\alpha\in\{x,y,z\}$.  One can also obtain higher moment correlations, e.g. $\langle (\hat n\cdot \v{S})^2 \rangle$, from the measurement record.  Between these pulses the spins evolve for a time $t$ under the Hamiltonian in Eq.~\eqref{eq:XXZ-Ham}.
We note that molecule experiments have recently begun
using this protocol~\cite{neyenhuis:anisotropic_2012} and Ref.~\cite{britton:engineered_2012} applied  it to long-range Ising models in recent Penning trap experiments with $\sim300$ ions.

If one imagines adding a transverse field term $h  \v{S}\cdot {\hat n}_\theta$ to the Hamiltonian in Eq.~\eqref{eq:XXZ-Ham}, with ${\hat n}_\theta$ a unit vector pointing $\theta$ from the $-z$ axis (see Fig.~\ref{fig:dynamic protocol}), the Ramsey protocol  corresponds to a quench from $h=\infty$ to $h=0$. One may therefore be able to explore, for example, Kibble-Zurek physics (e.g., entropy production, topological defects)~\cite{kibble_topology_1976,zurek_cosmological_1985}.

\textit{Theoretical methods.---}We calculate dynamics
in four limits: (1) short times, $\{J_\perp,J_z\}t \ll 1$, (2) Ising, $J_\perp=0$, (3) near-Heisenberg [SU(2)], $|J_z-J_\perp|\ll J_z$, and (4) one dimension for arbitrary $J_\perp/J_z$.  The first three limits' results are analytic and valid in any dimension, while the last is from numerically exact adaptive time-dependent density matrix renormalization group (adaptive t-DMRG)~\cite{white:density_1992,
vidal:efficient_2004,daley:time-dependent_2004,
schollwoeck:density-matrix_2005}. Details of the calculations will be presented elsewhere~\cite{hazzard:molecule_in-prep}.
In all cases $S^z$ is conserved, with $\expec{S^z(t)}=-(f/2)\cos\theta$.

\begin{figure}[t]
\setlength{\unitlength}{1.0in}
\includegraphics[width=3.4in,angle=0]{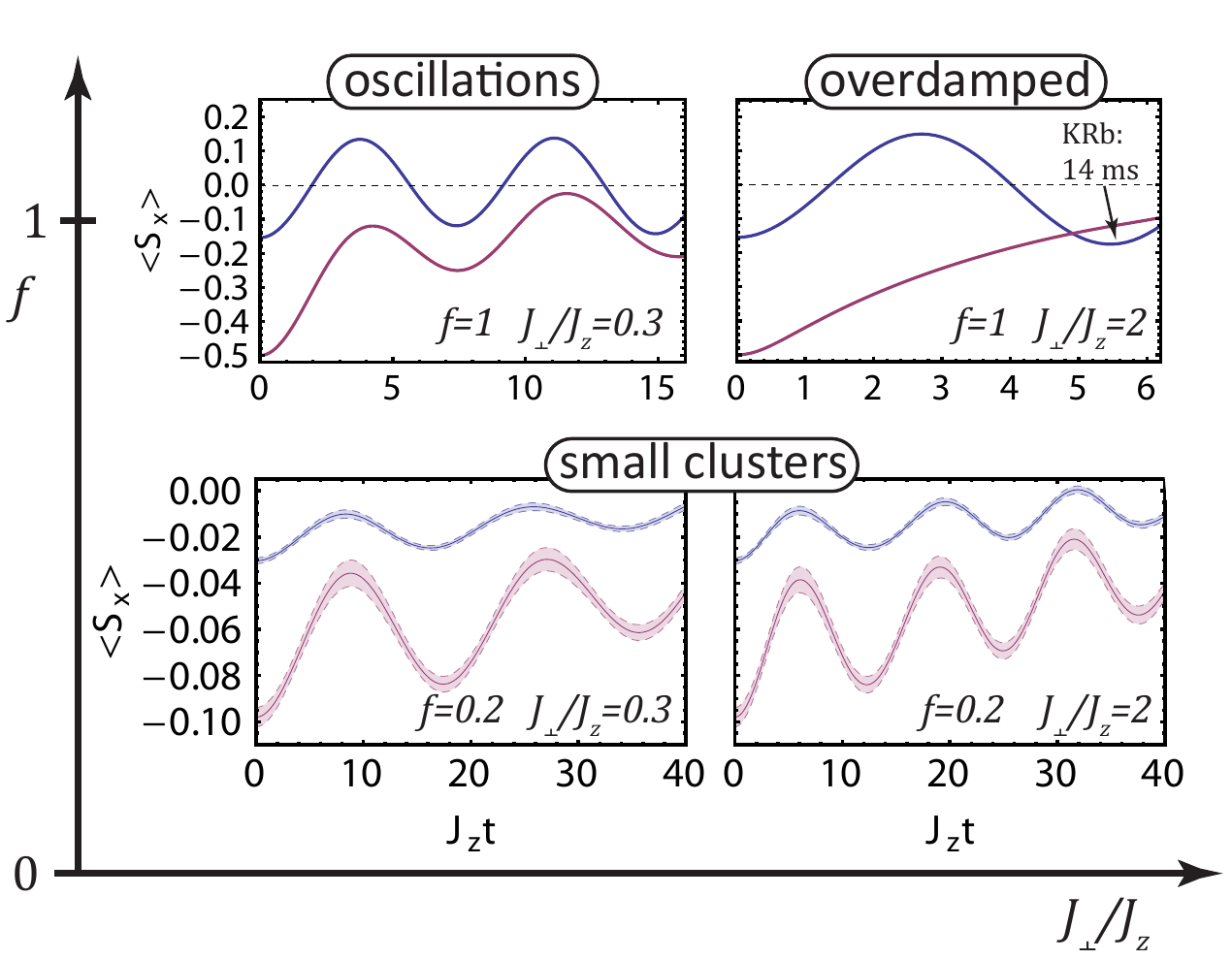}
\caption{ (Color online) ``Phase diagram" illustrating crossovers of dynamics versus filling $f$, $J_\perp/J_z$, and $\theta$ using t-DMRG on chains.  In each region, a plot shows  dynamics for $\theta=0.1\pi,0.5\pi$ (top blue, bottom purple), labeled with a qualitative description of the behavior.  Shaded regions in $f=0.2$ plots indicate one standard deviation errors from disorder averaging.  
``KRb: 14~ms" indicates the time for KRb in a one-dimensional 532~nm chain at a 5~kV/cm dc electric field giving $J_\perp/J_z=2$ for the two lowest energy, zero angular momentum projection rotational states.  Dynamics in higher dimensions is faster due to
having more neighbors.
 \label{fig:global-behavior}}
\end{figure}
\textit{Short time limit, $\{J_\perp,J_z\}t \ll 1$.}  For short times,   $\expec{\mc O(t)}= \expec{\mc O}-it\expec{[\mc O,H]}-\frac{t^2}{2} \expec{[[\mc O,H],H]} + O(t^3)$ for an operator $\mc O$.
We calculate the commutators and time dependence of  $\expec{S^\alpha(t)}$ to leading non-zero order, and  $\expec{S^\alpha(t)S^{\gamma}(t)}$ to linear order.  We find
\be
\expec{S_i^x} &=& \frac{f}{2}\sin \theta \left\{1-\frac{f\tau^2}{8}\lb \Xi_2 +f \Upsilon\cos^2\theta \rb\right\}
+O(\tau^4), \nonumber \\
\expec{S_i^y} &=& -(f^2 \tau \Xi_1/8)\sin(2\theta)  +O(\tau^3), \label{eq:short-time-means}
\ee
where $\tau=(J_z-J_\perp)t$, $\Xi_m=\sum_{j\ne0} V_{\text{dd}}^m(i,i+j)$, and $\Upsilon = \Xi_1^2-\Xi_2$.  Note that for these homogeneous systems, these observables are independent of $i$.  Similarly, defining ${\mc C}^{\alpha \gamma}_{ij} \equiv \expec{S^\alpha_i S^\gamma_j}$, we find
\be
{\mc C}^{xy}_{ij} &=& \frac{\tau f^3\sin(2\theta)\sin\theta }{16} \lb V_{\text{dd}}(i,j)-\Xi_1 \rb + O(\tau^2)  \label{eq:short-time-corrns}\\
{\mc C}^{yz}_{ij} &=& \frac{\tau f^3}{8}\lb\frac{ \sin(2\theta)\cos\theta }{2}  \Xi_1 
    +V_{\text{dd}}(i,j)\sin^3\theta \rb+ O(\tau^2)\nonumber
\ee
for $i\ne j$.
To linear order, ${\mc C}^{\alpha\alpha}_{ij}$ and  ${\mc C}^{xz}_{ij}$ are constant.
For $i=j$, the Pauli algebra reduces $\expec{S^\alpha_i S^\gamma_i}$ to $\expec{S_i^\delta}$.
One can compute $\Xi_m$  rapidly for arbitrary lattices and analytically in special cases (e.g. one dimension).

\begin{figure}[t!]
\setlength{\unitlength}{1.0in}
\includegraphics[width=3.45in,angle=0]{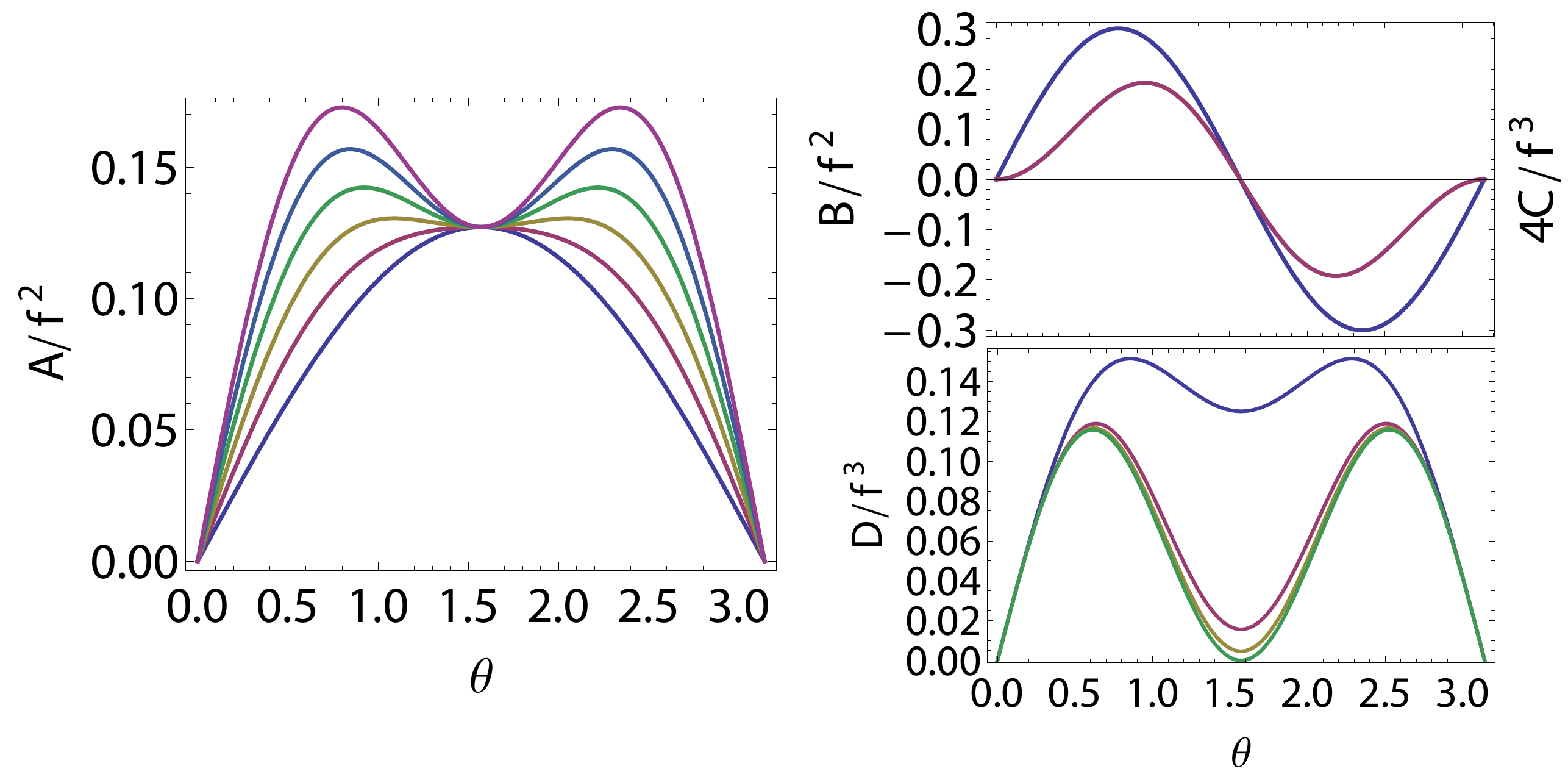}
\caption{ (Color online) Short time  behavior as functions of $\theta$ and $f$ in a one dimensional dipolar chain.  Left: coefficient $A/f^2$ defined by $\expec{S_i^x(t)}=\expec{S_i^x(0)}-A \tau^2+O(t^4)$ for fillings $f=0,0.2,\ldots,1.0$, bottom to top, with $\tau\equiv (J_z-J_\perp)t$. Right: $B$, $C$ (top, larger and smaller curves, respectively), and $D$ (bottom),
defined as
$\expec{S_i^y}=-B\tau$, $\expec{S_i^x S_j^y} = C (V_{\text{dd}}(i,j)-\Xi_1)\tau$ ($\Xi_1$ defined in text), and $\expec{S_i^yS_j^z}=D\tau$, respectively (to linear order). The coefficient $D$ is shown for $i-j=1,2,3,\infty$ (top to bottom). Results are from Eqs.~\eqref{eq:short-time-means} and~\eqref{eq:short-time-corrns}.
 \label{fig:short-time}}
\end{figure}

\textit{Ising limit, $J_\perp=0$.} We extend the Emch-Radin solution~\cite{emch:non-markovian_1966,radin:approach_1970,kastner:diverging_2011} for Ising dynamics to arbitrary $\theta$, inter-spin coupling strengths,  and to include disorder~\footnote{These results also extend straightforwardly to arbitrary spatially varying initial angles and longitudinal magnetic fields, as we will present elsewhere~\cite{hazzard:molecule_in-prep}.}.  We find
\be
\hspace{-0.15in}\expec{S_i^x(t)} \! &=&\! f\frac{\sin(\theta)}{2} \text{Re}\lb\prod_{j\ne i}\sum_{\rho_j}
g(\rho_j)e^{\frac{1}{2}it J_z V_{\text{dd}}(i,i+j)\rho_j}\rb\!\!,
\label{eq:Ising-1-site}
\ee
where the sum runs over $\rho=0$ (unoccupied site) and $\rho=\pm 1$ ($S^z=\pm 1/2$), and
\be
g(\rho) &= \begin{cases} 1-f \hspace{0.4in} & \text{if } \rho=0\\
f \sin^2(\theta/2) & \text{if } \rho=1 \\
f \cos^2(\theta/2) & \text{if } \rho=-1
\end{cases}.
\ee
The expectation $\expec{S_i^y}$ takes the imaginary (rather than real) part of the square-bracketed expression in Eq.~\eqref{eq:Ising-1-site}.
Similarly one can obtain correlations~\cite{hazzard:molecule_in-prep,kastner:in-prep}.
The product in Eq.~\eqref{eq:Ising-1-site} is readily evaluated numerically by truncating the interaction range, even for a truncation including thousands of sites.  In special limits $\expec{S_i^x}$ simplifies: e.g., for $\theta=\pi/2$  and $f=1$, $\expec{S_i^x(t)}=(1/2) \prod_{j\ne i} \cos(J_z V_{\text{dd}}(i,j)t/2)$.

\textit{Near-Heisenberg limit, $J_z\approx J_\perp$.---}Here a finite size gap $\Delta \propto J_\perp/N^2$ for $N$ particles
to excitations out of the Dicke manifold prevents states initially in the manifold from leaving it~\cite{rey:many-body_2008,rajapakse:single-photon_2009}.  Since matrix elements to leave it are $\propto J_z-J_\perp$, projecting to the Dicke manifold is accurate  when $|J_z-J_\perp|\ll \Delta$.
In this limit  the effective Hamiltonian is the collective spin $N/2$ model~\cite{rey:many-body_2008,rajapakse:single-photon_2009}
$
H_{\text{eff}} = \chi (S^z)^2 \label{eq:H-eff-near-SU2}
$
with
$
\chi = \frac{J_\perp-J_z}{N(N-1)}\sum_{i\ne j} V_{\text{dd}}(i,i+j).
$
Dynamics
are straightforwardly calculated for any disorder configuration, since there are only $N+1$ states in the Dicke manifold.  Here we restrict to $f=1$ for simplicity.
 For example, one finds
\be
\hspace{-0.2in}\expec{S^x_i} &=& \frac{\sin\theta}{2}\text{Re}\lb\lp \cos({\chi\tau})-i \cos\theta \sin({\chi\tau}) \rp^{{N}-1}\rb 
\label{eq:near-heisenberg-Sx-Sy}
\ee
Again $\expec{S_i^y}$ is the corresponding imaginary part.
Unlike the other approximations, this is valid only for finite  $N$.

\textit{One dimension.---}We use adaptive t-DMRG~\cite{white:density_1992,
vidal:efficient_2004,daley:time-dependent_2004,
schollwoeck:density-matrix_2005} to calculate dynamics of one dimensional chains.  We treat $20$ site chains and find finite size effects to be fairly small.  We discretize time in steps of  $0.05J_z^{-1}$,  and find a discarded weight of $\lsim 10^{-9}$ for times $\lsim 10J_z^{-1}$, adaptively keeping $m=50$-$500$ reduced density matrix states.  
Altogether, we expect errors dominated by the disorder average, which is taken over 100 random configurations.

\begin{figure}[t]
\setlength{\unitlength}{1.0in}
\includegraphics[width=3.35in,angle=0]{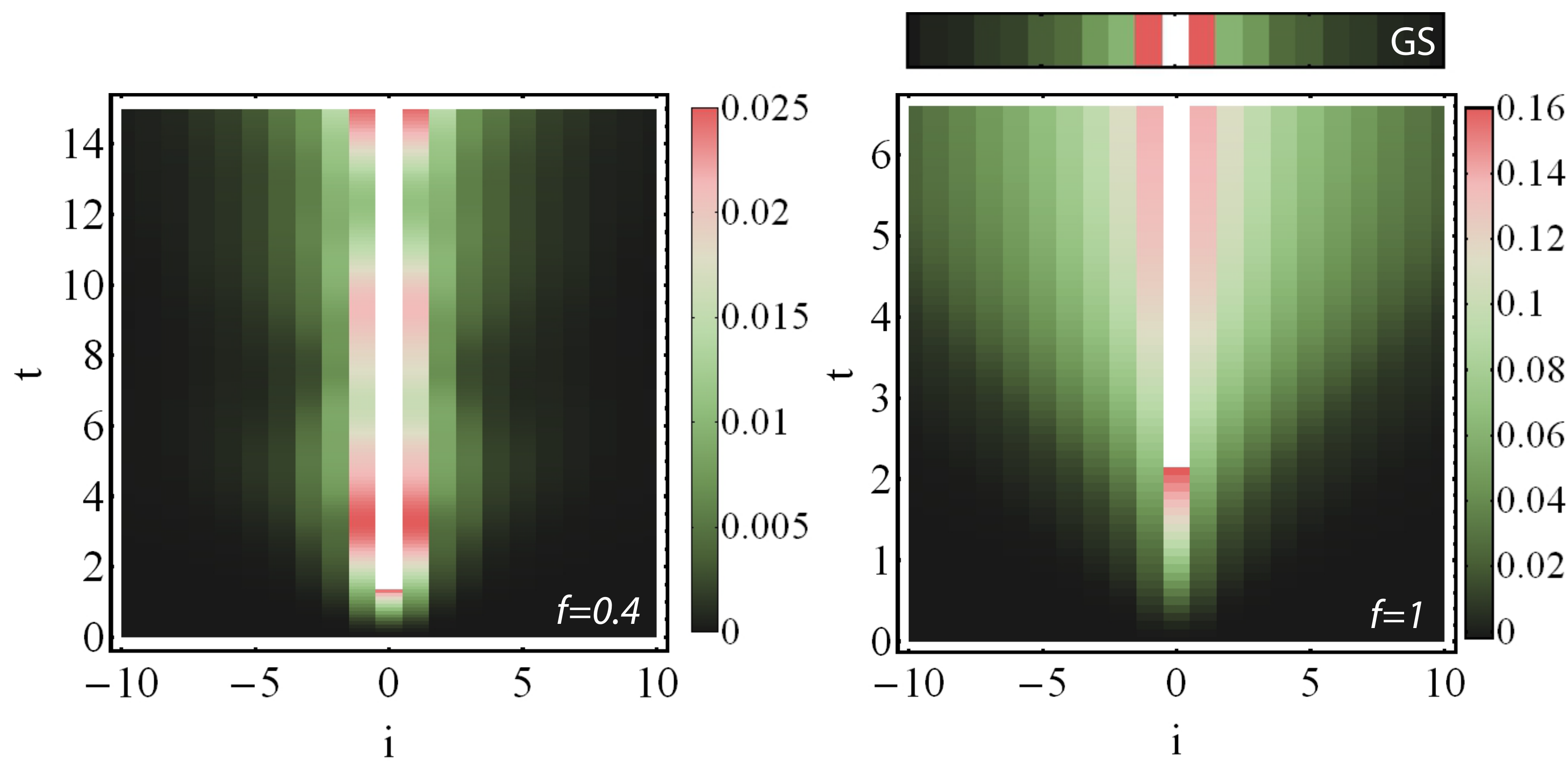}
\caption{ (Color online) Correlation dynamics and comparison with ground state: $\expec{S_j^x S_{j+i}^x}-\expec{S_j^x}\expec{S_{j+i}^x}$ (averaged over $j$) for $J_\perp=2J_z$ with $\theta=\pi/2$ as a function of time for   $f=0.4$ (left) and $f=1$ (right), compared to the ground state (upper right bar).  Other $f$, $\theta$, and $J_\perp$ are similar.
 \label{fig:develop-correlation}}
\end{figure}

\textit{Results: global perspective.---}Fig.~\ref{fig:global-behavior} overviews dynamics,  from the calculations above,
 as a function of $J_\perp/J_z$,   $f$, and  $\theta$.  Experimentally, these are controlled by  electric field~\cite{gorshkov:quantum_2011,gorshkov:tunable_2011},  temperature/density, and  first Ramsey pulse area, respectively.  Fig.~\ref{fig:global-behavior} shows dimension $d=1$ results, but  our analytic expressions show that the $d=1$ results are representative of  $d>1$.  Dynamics in $d>1$ have more neighbors and thus is faster.

Consider  $f=1$ and $\theta=\pi/2$.   For $J_\perp/J_z=0$, $\expec{S_i^x}$ oscillates with period $2\pi/J_z$ from the nearest neighbor interaction, superposed with slower oscillations  from longer range interactions.  The first-few-neighbor interactions  account for the dynamics to times $t\sim 10 J_z^{-1}$.  For $J_\perp=0$ the frequencies  form a discrete set.
Increasing $J_\perp$ gives a continuum of frequencies, damping the oscillations. Approaching $J_\perp=J_z$, the dynamics slows down, since at $J_\perp=J_z$ the initial state is an eigenstate of the Hamiltonian.  As $J_\perp/J_z$  increases further, the dynamics is damped with characteristic timescale $(J_\perp-J_z)^{-1}$.

For  $f\ll 1$, the behavior crosses over to that of  independent clusters, eventually with only two particles.
The largest frequency is roughly half that for $f=1$, since there is a single neighbor instead of two. Thus, the dynamics remains roughly as fast as for $f=1$, but  the dynamics' magnitude at times $\sim \{J_z^{-1},J_\perp^{-1}\}$ is smaller since there are fewer molecules and only a fraction of them are close enough to interact.
At any $f$ the overall timescales are roughly independent of  $\theta$ but the damping vanishes as $\theta\rightarrow0$.

\textit{Achieving goals of emulating quantum magnetism.---}Fig.~\ref{eq:short-time-corrns} shows the characteristic dependence of the XXZ model's short time dynamics on $\theta$, $f$, and $\{J_z,J_\perp\}$.  This can be used to experimentally verify the emulation of the XXZ model and benchmark its accuracy.


To achieve the second goal of generating interesting well-understood states, both the near-Heisenberg and Ising limits are useful.  For $f\approx 1$ and $\theta=\pi/2$ near the Heisenberg point, the state at $t=\pi/(2\chi)$  is $\ket{GHZ}=(1/\sqrt{2})(\ket{\leftarrow\cdots \leftarrow}+e^{i\phi}\ket{\rightarrow\cdots \rightarrow})$ for some  $\phi$~\cite{molmer:multiparticle_1999,wang:spin_2001,rey:many-body_2008,rajapakse:single-photon_2009}.  This is a cat state, specifically the GHZ or NOON state, which is useful for metrology~\cite{bollinger:optimal_1996}.
 Ising dynamics offer other interesting states. For nearest neighbor interactions and $\theta=\pi/2$, the state at $t=\pi/(2J_z)$ is a cluster state, which suffices for universal measurement based quantum computing~\cite{raussendorf:one-way_2001}.  The presence of long range interactions perturbs the cluster state, and an interesting question is how this affects its utility.  Decoherence  can also limit the creation of entangled states.

A generic implementation of the proposed dynamics in $d>1$  achieves the third goal, emulating quantum magnetism in theoretically intractable regimes.  Away from the short time, Ising, and Heisenberg limits, no solution is known in $d>1$.  As Fig.~\ref{fig:develop-correlation} shows, in $d=1$ strong correlations develop, suggesting the difficulty of $d>1$ calculations.  The  long time $f=1$ correlations  are even larger at large distance than  in the ground state.  Interestingly, the dynamics shows a light-cone-like spreading to an apparent steady state.

\textit{Experimental outlook.}---Though our discussion focused on molecules, we point out that the dynamics studied here can have direct application in other physical systems,  including condensed matter~\cite{quilliam:experimental_2012}, trapped ions~\cite{kim:quantum_2010,britton:engineered_2012}, and  optical lattice clocks~\cite{swallows:suppression_2011,lemke:p-wave_2011}.

We close by noting technical details for molecule experiments.    Rotational states' polarizabilities differ~\cite{kotochigova:electric_2010,gorshkov:tunable_2011}, so the optical trap induces a spatially varying field $\sum_i h_i S_i^z$. Also, Eq.~\eqref{eq:XXZ-Ham} ignores density-density
$n_i n_j$ and density-spin $n_i S_j^z$ interactions~\cite{gorshkov:tunable_2011,gorshkov:quantum_2011}.  For $f<1$, the latter gives a spatially varying magnetic field that depends on molecules' random positions.
Spin-echo pulses common in Ramsey experiments remove both effects.

\textit{Summary.---}We have shown that Ramsey spectroscopy enables ongoing ultracold polar molecule experiments to accomplish three goals for emulating quantum magnetism: (1) benchmarking the emulation's accuracy (using short time dynamics), (2) generating strongly correlated and entangled states in well-understood limits (Ising, near-Heisenberg, one dimension), and (3) exploring strongly correlated dynamics in regimes inaccessible to theory (generic case in dimensions $d>1$).

Finally, we mention that in addition to the XXZ Hamiltonian explored in this paper, our dynamic protocol should be useful for verifying emulation of
more complicated spin models that may be realized with ultracold molecules, as in Refs.~\cite{gorshkov:tunable_2011,gorshkov:quantum_2011} and beyond.

\textit{Acknowledgements.---}We thank A. Gorshkov, J. Bollinger, M. Kastner, M. Lukin, J. Ye, D. Jin, and the Jin-Ye molecule group for numerous  conversations.  KH thanks the NRC for support and the Aspen Center for
Physics, which is supported by the NSF, for its hospitality during the initial conception of this work. This work utilized the Janus supercomputer, which is supported by the NSF (award number CNS-0821794) and the University of Colorado Boulder, and is a joint effort with the University of Colorado Denver and the National Center for Atmospheric Research. AMR acknowledges support from the NSF (PFC and PIF), ARO individual investigator award and ARO with funding for the DARPA-OLE program.

\bibliography{molecules-far-from-eq-short-v1}

\end{document}